\documentclass[angeo,manuscript]{copernicus}

\usepackage{soul}

\begin{document}

\nolinenumbers

\title{Comment on ``Cavitons and spontaneous hot flow anomalies in a hybrid-Vlasov global magnetospheric simulation'' by Blanco-Cano et al. (2018)}

\Author[1]{G{\'a}bor}{Facsk{\'o}}

\affil[1]{Rhea System GmbH., TIZ Building, Robert Bosch Str. 7., 64293 Darmstadt, Germany}

\runningtitle{Comments on ''Cavitons and SHFAs in a hybrid-Vlasov global magnetospheric simulation''}

\runningauthor{G.~Facsk{\'o}}

\correspondence{G{\'a}bor Facsk{\'o} (g.facsko@rheagroup.com)}

\received{\today}
\pubdiscuss{\today} %% only important for two-stage journals
\revised{\today}
\accepted{\today}
\published{\today}

%% These dates will be inserted by Copernicus Publications during the typesetting process.

\firstpage{1}

\maketitle

\copyrightstatement{The Author}

%% Paper

\citet{blanco-cano18:_cavit_vlasov} analysed the output of the VLASIATOR global hybrid Vlasov solver and intended to find Spontaneous Hot Flow Anomalies \citep[SHFA,][]{zhang13:_spont,omidi13:_spont}. This is a very nice paper about the development of the foreshock cavitons and magnetosheath cavities based on unique global hybrid-Vlasov simulations. However, the simulation results cannot reproduce the main features of the SHFAs. The SHFAs (and the HFAs) show density and magnetic field magnitude drops in their cavity. The magnetic field is turbulent in the cavity. The temperature is very high in them, a few million K. The solar wind direction turns away from the radial direction and slows down \citep{facsko10:_study_clust,zhang13:_spont,omidi13:_spont}. The latter features gave the name of the phenomenon hence I had serious concerns about whether the Authors had detected SHFA in the paper above.

On Figure~3,~7,~9 of the paper above, the Authors see density and magnetic field drops. The simulated phenomena is not significantly hotter than the surrounding foreshock plasma. The foreshock plasma temperature is never observed at 10\,million\,K. Hence locating in the foreshock cannot be an excuse for the missing feature of the phenomena. The Authors also see “\textit{[\dots] deviations from the bulk solar wind velocity are observed throughout the foreshock, and they are not prominent enough inside SHFAs to be unambiguously identified.}” The phenomenon that does not show anomalous flow cannot be called Spontaneous Hot Flow Anomaly.

The SHFAs are surrounded by density and magnetic field increasement at the edge of the phenomena. Their presence proves that the cavity is not in equilibrium and expands. The hybrid simulations of \citet{omidi13:_spont} could present these shoulders (\citet{lin02:_global} could also have simulated them for HFAs). Furthermore, these increasements lead to the observed depletion of the solar wind velocity because the deceleration of the solar wind comes from the bad fitting and plasma moment calculation \citep[Figure 3, 7]{parks13:_reint_slowd_solar_wind_mean, kecskemety06:_distr_rapid_clust}. Hence, it is possible to explain the missing solar wind deceleration if these increasements are present. If both features are missing, the phenomena cannot be SHFA.

The Authors also study foreshock cavitons, magnetosheath filaments and structures in this paper above. My comments are limited only to the identification and analysis of the ``SHFA events'' of the simulation. Based on the remarks described above, I am sure that the features in the VLASIATOR simulations are not SHFAs. However, these questionable events could develop to an SHFA. \citet{zhang13:_spont} observed SHFA--like events without significant solar wind deceleration. As \citet{zhang10:_time_histor_event_macros_inter_subst} discovered and introduced the phenomenon of so called proto-HFA, \citet{zhang13:_spont} discovered the phenomena of proto-SHFA. These proto-SHFAs were simulated by the VLASIATOR code and misintepreted by the Authors.

\begin{acknowledgements}
G{\'a}bor Facsk{\'o} thanks Sophie Burley for improving the English of the paper.
\end{acknowledgements}

%% REFERENCES

\bibliographystyle{copernicus}
%\bibliography{angeo-2019-6-tx.bib}

%%
%% URLs and DOIs can be entered in your BibTeX file as:
%%
%% URL = {http://www.xyz.org/~jones/idx_g.htm}
%% DOI = {10.5194/xyz}

%% LITERATURE CITATIONS
%%
%% command                        & example result
%% \citet{jones90}|               & Jones et al. (1990)
%% \citep{jones90}|               & (Jones et al., 1990)
%% \citep{jones90,jones93}|       & (Jones et al., 1990, 1993)
%% \citep[p.~32]{jones90}|        & (Jones et al., 1990, p.~32)
%% \citep[e.g.,][]{jones90}|      & (e.g., Jones et al., 1990)
%% \citep[e.g.,][p.~32]{jones90}| & (e.g., Jones et al., 1990, p.~32)
%% \citeauthor{jones90}|          & Jones et al.
%% \citeyear{jones90}|            & 1990

%%% PHYSICAL UNITS
%%%
%%% Please use \unit{} and apply the exponential notation

\end{document}